# Design of Optimal Topology of Satellite-Based Terrestrial Communication Networks

Boris S. Verkhovsky

**Abstract**-Topological design of terrestrial networks for communication via satellites is studied in the paper. Quantitative model of the network cost-analysis minimizing the total transmission and switching cost is described. Several algorithms solving combinatorial problem of the optimal topology design based on binary partitioning, a minimax parametric search and dynamic programming are developed by the author and demonstrated with a numeric example. Analysis of average complexity of the minimax parametric search algorithm is also provided.

**Index terms**-satellite communication network, terrestrial networks, network topology design, switching/transmission cost, network-cost analysis, binary partitioning, dynamic programming, average complexity, clustering, combinatorial problem

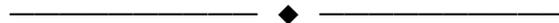

## 1. INTRODUCTION

Modern wide-area satellite communication networks consist of terrestrial users interconnected via terrestrial links with routers/switches called earth stations (ES). An earth station (ES) communicates as a transmitter and receiver with one or several satellites [6], [11], [20], [21], [25]. Widely dispersed "satellite dishes" do not provide quality two-way communications. Only large corporations, major governmental agencies, and large telecommunications vendors can afford individual ESs. Small or medium sized corporations among other users must share a single ES.

Modern telecommunications is a highly competitive business that strives to reduce service fees to increase market share by making their services more economically attractive to potential customers. [1]Such a communications company must expertly locate its various ESs, which may be of different capacities, and also decide how its customers should be interconnected with these ESs, [1], [3], [5], [8], [18], [19], [26], [27]. An optimally designed network can potentially save hundreds of millions of dollars annually and thereby attract additional users with its lower service fees [10].

From a computational point of view, the network design task is a formidable combinatorial problem, i.e., it requires brute-force algorithms or heuristics with exponential time-complexity, because they must determine an optimal way of clustering all users, [7], [12], [14].

Several algorithms developed by the author [22]-[24] are described in this paper and demonstrated with a detailed numeric example. These algorithms are based on statistical properties observed by the author in thousands of computer experiments. They solve the problem of clustering and locating all ESs with a polynomial time complexity. All related proofs are provided in [22]-[24]. For additional insights into the problems and algorithms related to network design see [2], [10], [15]-[17], [21].

## 2. PROBLEM STATEMENT

1). Let us consider the locations of $n$ users with coordinates $P_i=(a_i,b_i)$, $i=1,…,n$. Each user is characterized by a "volume" of incoming and outgoing communication flow $w_i$ ("weight" of the $i$-th user);

---

[1] *Boris S. Verkhovsky is with Computer Science Department, New Jersey Institute of Technology; Newark, NJ 07102, USA.*

2). Let $C_k = (u_k, v_k)$ denote the location of the $k$-th ES, $k=1,2,..,m$;

3). Let $S_k$ be a set of all users connected with $C_k$;

4). Let $f(w_i, P_i, C_k)$ describe a cost function of the transmission link connecting $P_i$-user and $C_k$.

For all $i=1,2,..,n$ $P_i$ are the inputs and for all $k=1,2,..,m$ $C_k$ and $S_k$ are the decision variables/outputs.

With these inputs, a minimal total cost of all terrestrial links and all ESs equals

$$\min_{S_1,...,S_m} \min_{C_1,...,C_m} \left[ \sum_{k=1}^{m} \sum_{i \in S_k} f(w_i, P_i, C_k) + q_k\left(\sum_{i \in S_k} w_i\right) \right] \quad (2.1)$$

where $q_k\left(\sum_{i \in S_k} w_i\right)$ is the cost of $k$-th ES representing a non-linear function of all outgoing and incoming flows. Thus the problem (2.1) requires a comprehensive analysis to determine the optimal clusters (subsets) $S_1,..,S_m$ and locations of the routers/ESs $C_1,..,C_m$.

Complexity of clustering in general has been studied and described in [7]. Surveys on quantitative modeling and algorithms related to clustering are provided in [12] and [14].

## 3. FOUR SPECIAL CASES

**Case1:** If the locations of all switches/ESs are specified and the cost function of every ES is flow-independent, then it is easy to find the clusters $S_k$. Indeed,

$$S_k := \left\{ i : \min_{1 \le j \le m} f(w_i, P_i, C_j) = f(w_i, P_i, C_k) \right\} \quad (3.1)$$

**Case2:** If for $k=1,…,m$ $S_k$ are known, then the optimal location of every ES can be determined independently:

$$\min_{C_k} \sum_{i \in S_k} f(w_i, P_i, C_k) \text{ for } k=1,…,m. \quad (3.2)$$

**Case3:** If $f(w_i, P_i, C_k) = w_i \, dist(P_i, C_k)$ (3.3)

then the problem (3) is known as a *Weber problem*. This class of problems has been investigated by many authors over the last forty years, [4] and [9].

**Case4:** If $q_k\left(\sum_{i \in S_k} w_i\right)$ is a linear or convex function, i.e.,

$$q_k(w^1 + w^2) \ge q_k(w^1) + q_k(w^2), \quad (3.4)$$

then

$$q_k\left(\sum_{i \in S_k} w_i\right) \ge q_{2k}\left(\sum_{i \in S_{2k}} w_i\right) + q_{2k+1}\left(\sum_{i \in S_{2k+1}} w_i\right),$$

which implies that the greater the number of clusters the lower the total costs of all routers/ESs.

Difficulties arise if

- the clusters $S_k$ are not known; or
- the cost of every ES is neither small nor flow-independent; or
- the number $m$ of ESs **and** their optimal locations ($u_k, v_k$) for every $k$ are not known.

## 4. PARTITION INTO TWO CLUSTERS

It is important to stress that there is a substantial difference between the two cases: $m=1$ and $m=2$. In the latter case the problem can be solved by repetitive application of an algorithm designed for the Weber problem. This must be done for all possible pairs of clusters $S_1$ and $S_2$. There are $2^{n-1}-1$ different ways to partition $n$ points into two subsets $S_1$ and $S_2$ and, for each clustering, two Weber problems must be solved. Thus, even for $m=2$ the total time complexity of this brute-force combinatorial approach is O($2^n$), [13].

## 5. BINARY PARAMETRIC PARTITIONING

In this section we provide a procedure that divides a network $N$ with one ES $S$ into two sub-networks $N_1$ and $N_2$ with two earth stations and *two clusters* $S_1$ and $S_2$.

**StepA1**: {find an optimal location of the "center of gravity" $C_0$ for all $n$ users, [18]}: consider $m=1$ and solve the

Problem $\min_C \sum_{i=1}^{n} f(w_i, P_i, C)$; (5.1)

**StepA2**: Consider a straight line $L$ and rotate it around the center of gravity {CoG} $C_0$;

**StepA3**: For every user consider their polar coordinates $(d_i, \varphi_i)$ using $C_0$ as the origin of the coordinate system; {here $\varphi_i$ is an angular coordinate of $P_i$};
*Remark*1: The line $L$ divides all $n$ points into two clusters, $S_1(x)$ and $S_2(x)$, by at most $n$ different ways as the angle $x$ increases from 0 to $\pi$;
**StepA4**: **for** $i$=1 **to** $n$ **do**
**if** $\pi \leq \varphi_i < 2\pi$, **then** $x_i := \varphi_i - \pi$; (5.2)
  **sort** all $x_i$ in ascending order;
**StepA5**: **if** $x_i = \varphi_i$, **then** $c_i := 1$ **else** $c_i := -1$;
**StepA6**: **if** $(x - x_i)c_i \geq 0$, (5.3)
  **then** $P_i \in S_1(x)$ **else** $P_i \in S_2(x)$;
{see Table1 for illustration};

**Table1**: {using, as example, $x$=1.53}

| $\varphi_i$ | $x_i$ | $c_i$ | $P_i(x) \in S_k$ |
|---|---|---|---|
| 3.65 | .51 | -1 | $P_1(x) \in S_2$ |
| .67 | .67 | 1 | $P_2(x) \in S_1$ |
| 1.53 | 1.53 | 1 | $P_3(x) \in S_1$ |
| 2.11 | 2.11 | 1 | $P_4(x) \in S_2$ |

**StepA7**: **for** $k$=1,2 and $P_i \in S_k(x)$ (5.4)
compute
$$g_k(S_k(x)) := \min_{C_k} \sum_{i \in S_k} f(w_i, P_i, C_k); \quad (5.5)$$

**StepA8**: {compute the cost of two routers/ESs and all connecting links}:
$$h(x) := \sum_{j=1}^{2} \left[ q_j \left( \sum_{i \in S_j} w_i + g_j(S_j(x)) \right) \right] \quad (5.6)$$

**StepA9**: {rotate the line $L$ and find an angle that minimizes function $h(x)$}:
$$h(r) := \min_{0 \leq x \leq \pi} h(x); \quad (5.7)$$

**StepA10**: **if** for $i \in S_1(r)$
  $f(w_i, P_i, C_1) > f(w_i, P_i, C_2)$,
    **then** reassign $i \in S_2(r)$; (5.8)
**if** for $i \in S_2(r)$ $f(w_i, P_i, C_2) > f(w_i, P_i, C_1)$,
    **then** reassign $i \in S_1(r)$; (5.9)

**StepA11**: using (5.8) and (5.9), update optimal locations of $C_1$ and $C_2$ for new values of $S_1(r)$ and $S_2(r)$;
*Remark*2: we define $S_1(r)$ and $S_2(r)$ as the *optimal binary partitioning*.

## 6. SEARCH FOR THE "CENTER OF GRAVITY"

**StepB1**: assign *flag*:=0;
$$u := \sum_{i \in N_1} w_i a_i / \sum_{i \in N_1} w_i; \\ v := \sum_{i \in N_1} w_i b_i / \sum_{i \in N_1} w_i; \quad (6.1)$$

**StepB2**: compute for every $i \in N_1$
$$R_i := \sqrt{(u - a_i)^2 + (v - b_i)^2}; \quad (6.2)$$

**StepB3**: $old(u,v) := (u,v)$; compute
$$u := \left( \sum_i w_i x_i / R_i \right) / \left( \sum_i w_i / R_i \right); \\ v := \left( \sum_i w_i y_i / R_i \right) / \left( \sum_i w_i / R_i \right); \quad (6.3)$$

**StepB4**: **while** $dist\left[ old(u,v), (u,v) \right] > \varepsilon$
**repeat** Steps B2 and B3; {search for a stationary point $SP$; $\varepsilon$ is a specified accuracy for the location of the CoG};
**StepB5**: let $SP := (u,v)$;
**StepB6**: **if** for all $j \in N_1$ $dist(SP, P_j) > \varepsilon$
and *flag*=0, **then** $SP$ is the CoG;
**if** for all $j \in N_1$ $dist(SP, P_j) > \varepsilon$; (6.4)
and $flag = -1$,
**then** $N_1 := N_1 + \{pnt\}$; *flag*:=0; (6.5)
**goto** StepB2;
**StepB7**: **if** $dist(SP, P_k) \leq \varepsilon$, (6.6)
**then** $flag = -1$; $pnt := k$; $N_1 := N_1 - \{pnt\}$.

For validation of the *CoG* algorithm see Lemmas 1 and 2 in the Appendix;
*Remark*3: Table2 lists all possible cases of the algorithm:

**Table2**

|  | For $\forall j: dist(SP, P_j) > \varepsilon$ | $dist(SP, P_k) \leq \varepsilon$ |
|---|---|---|
| $flag=0$ | SP is "center of gravity" | $pnt:=k$; $oldSP:=P_k$; $N_1:=N_1-\{pnt\}$; $flag:=-1$ |
| $flag=-1$ | if $oldSP := P_k$, then SP is "center of gravity" | $N_1 := N_1 + \{pnt\}$; do local search for a minimum of function (5.1) |

## 7. MINIMAX SEARCH FOR min$h(x)$

Let $h$ be a function computable on a set $S$ of $M$ discrete points $x_1,...,x_M$. We demonstrate an optimal search algorithm designed for the case where $h$ is a periodic function with known period $P$, i.e., $h(x_i + sP) = h(x_i)$ holds for every integer $s$ and for every $i=1,..,M$. Here all values $x_1,...,x_M$ are known.

It is obvious that $M$ evaluations of $h$ at points $x_1,...,x_M$ are sufficient to solve any problem by total enumeration. The optimal search algorithm provided below requires time complexity of order $\Theta(\log M)$.  (7.1)

For the sake of simplicity of notation let
$$h_i := h(x_i) \text{ and } g(t) := h(x_t). \quad (7.2)$$

Below we provide the optimal search algorithm for the case if $M = F_n$, where $F_n$ is $n$-th Fibonacci number.

The algorithm can be adjusted if
$$F_{n-1} < M < F_n. \quad (7.3)$$

A detailed description of the algorithm searching for minimum of a function and proof of its optimality are provided in [22], [23].

## 8. OPTIMAL SEARCHING ALGORITHM

The algorithm is optimal in the following sense: Let $H$ be a set of all functions of a period $P$; let $Q$ be a set of all possible strategies that find a minimum $h_r$ of $h(x)$; and let $e(h,q)$ be the number of required evaluations of $h(x)$ to determine the minimum $h_r$. Then $q^*$ is optimal in the worst case if

$$\min_{q \in Q} \max_{h \in H} e(h,q) = \max_{h \in H} e(h, q^*). \quad (8.1)$$

**StepC1**: if $F_n = 1$, then $h_r := h_1$; $x_r := x_1$; **stop**; **else** select a random integer $L_0$;
$$R_0 := L_0 + F_{n-1}; \quad (8.2)$$

**StepC2**: compute $g(R_0)$ and $g(L_0)$;

**StepC3**: {selecting an initial detecting state}; **if** $g(L_0) \geq g(R_0)$, **then**
$$\begin{aligned} A_1 &:= L_0; B_1 := L_0 + F_n; \\ R_1 &:= R_0; L_1 := A_1 + F_{n-2}; \end{aligned} \quad (8.3)$$

**else**
$$\begin{aligned} B_1 &:= R_0; A_1 := R_0 - F_n; \\ L_1 &:= L_0; R_1 := B_1 - F_{n-2}; \end{aligned} \quad (8.4)$$

**StepC4**: **if** $g(L_k) \geq g(R_k)$; **then**
$$\begin{aligned} A_{k+1} &:= L_k; temp := g(R_k); \\ L_{k+1} &:= R_k; R_{k+1} := 2L_k - A_k; \end{aligned} \quad (8.5)$$

compute $g(R_{k+1})$; assign
$$B_{k+1}:=B_k; I_k:=B_k-L_k; g(L_{k+1}):= temp; \quad (8.6)$$

**else** assign
$$\begin{aligned} B_{k+1} &:= R_k; temp := g(L_k); \\ R_{k+1} &:= L_k; L_{k+1} := 2R_k - B_k; \end{aligned} \quad (8.7)$$

compute $g(L_{k+1})$; assign
$$A_{k+1}:=A_k; I_k:=R_k-A_k; g(R_{k+1}):= temp; \quad (8.8)$$

*Remark*4: $I_k$ is the size of the interval of uncertainty containing a minimizer of $h(x)$ after $k$ evaluation of this function;

**StepC5**: **while** $I_k > 1$ **repeat** StepC4;

**StepC6**: $h_r := temp$; **stop**.

## 9. BINARY PARTITIONING AND ASSOCIATED BINARY TREE

Let us consider an algorithm that divides the network/cluster $N_1$ into two subnetworks $N_2$ and $N_3$ with corresponding transmission costs $t_2$ and $t_3$ and corresponding costs of ESs $q_2$ and $q_3$. Let $$d_k := t_k + q_k, \quad (9.1)$$
where $d_k$ is the hardware cost of the network $N_k$.

We assume that the algorithm divides $N_1$ into two subnetworks in such a way that $d_2 + d_3$ is minimal. For further consideration we represent the binary partitioning as a binary tree where the root of the tree represents a cluster (set of all users) $S_1$ and associated with it network $N_1$. In general, a $k$-th node of the binary tree represents a cluster $S_k$ and associated with it sub-network $N_k$. The two children of the $k$-th node represent two subnetworks $N_{2k}$ and $N_{2k+1}$ that resulted from the binary partitioning of the network $N_k$.

From the above definitions and from the essence of the problem it is clear that for all $k$ the following inequalities hold:

$q_k \geq q_{2k}$, $q_k \geq q_{2k+1}$ and $t_k \geq t_{2k} + t_{2k+1}$. (9.2)

The latter inequality holds because each sub-network $N_{2k}$ and $N_{2k+1}$ has a smaller number of users than $N_k$.

## 10. ANALYSIS OF HARDWARE COST

If $\quad d_k > d_{2k} + d_{2k+1}, \quad (10.1)$
then it is obvious that a partitioning into two clusters (subnetworks) is cost-wise beneficial. Yet, $d_k < d_{2k} + d_{2k+1}$ does *not* imply that any further partitioning is not cost-wise beneficial. To illustrate this let us consider a network $N_k$ and its *six* sub-networks $N_{2k}, N_{2k+1}, N_{4k}, N_{4k+1}, N_{4k+2}, N_{4k+3}$.

*Remark*5: To demonstrate various cases we consider two scenarios with inputs for $t_k$ and for $t_{4k+1}$ as shown in Table3:

A) $t_k = 34$ and for $t_{4k+1} = 5$;

B) $t_k = 31$ and for $t_{4k+1} = 3$.

**Case A:** $d_k = 46$; since $d_k > d_{2k} + d_{2k+1}$, (10.2)
then the binary partitioning of $N_k$ into two subnetworks is worthwhile;

**Case B:** $d_k = 43$; {local costs-analysis of hardware does not provide a correct insight}; in this case $d_k < d_{2k} + d_{2k+1}$, (10.3)
which only implies that there is no reason to divide the network $N_k$ into two subnetworks $N_{2k}$ and $N_{2k+1}$.

However, further analysis shows that
$\quad d_k > d_{4k} + d_{4k+1} + d_{4k+2} + d_{4k+3}$ (10.4)
if $\quad d_{4k+1} = 10$;
and $\quad d_k > d_{2k} + d_{4k+2} + d_{4k+3}$ (10.5)
if $\quad d_{4k+1} = 12$.

These examples illustrate that for a proper partitioning a *global* rather than a local analysis is required.

*Definition*1: We say that a network $N_k$ is *indivisible* if there is no cost-wise advantage to dividing it any further.

In addition, a network designer may stipulate that some sub-networks may not be further divisible if they do not satisfy at least one of the following threshold conditions:

a) Their combined "weight" {incoming and outgoing flow $w$} is lower than a specified threshold;

b) The number of users in the cluster {sub-network} is smaller than a specified by the designer threshold.

*Definition*2: We say that an optimal configuration of a communication network is determined if all indivisible sub-networks of the initial network $N_1$ are known.

## 11. DYNAMIC PROGRAMMING ALGORITHM

This algorithm initially assigns labels to all nodes of the associated binary tree, then determines final labels and then

**Table3**

| Subnetworks $N_i$ | $N_k$ | $N_{2k}$ | $N_{2k+1}$ | $N_{4k}$ | $N_{4k+1}$ | $N_{4k+2}$ | $N_{4k+3}$ |
|---|---|---|---|---|---|---|---|
| Transmission cost $t_i$ | 34; 31 | 13 | 15 | 6 | 5; 3 | 5 | 4 |
| ES cost $q_i$ | 12 | 9 | 8 | 5 | 7 | 6 | 4 |
| Hardware cost $d_i$ | 46; 43 | 22 | 23 | 11 | 12; 10 | 11 | 8 |

finds the optimal clustering. It consists of two stages: bottom-up stage and top-down stage. The algorithm described below was developed by the author of this paper years ago, but it has not been published.

### 11.1 Assignment of final labels
Here we assume that for all *k*=1, 2,…,*m* the values of all $d_k$ are pre-computed.

*Bottom-up stage*:
a) Assign to *k*-node a label:
$$L_k := d_k, k=1, 2,…,m;  \quad (11.1)$$
b) **if** *i*-th node is a leaf, **then** its final label
$$F_i := L_i; \quad (11.2)$$
c) **if** both children of *k*-th node have final labels, **then**
$$F_k := \min(L_k, F_{2k}+F_{2k+1}); \quad (11.3)$$
d) **if** the final labels $F_k$ are computed for all nodes, **then goto** the next stage.

For explanations see paragraph 12.2 and Fig.1-3 in the Appendix.

### 11.2 Principle of optimality
*Top-down stage*:
e) Starting from *j*=1 assign for every node, **if** $L_j = F_j$,
   **then** $w_j := 1$ **else** $w_j := 0$ (11.4)
g) {*Principle of optimality*}: if for the *k*-th node $w_k := 1$ and for every its ancestor $a(k)$ $w_{a(k)} := 0$, then this node is optimal and the corresponding cluster $S_k$ is non-divisible. Therefore, the sub-network $N_k$ is optimal.

*Remark*6: It can be shown that it is not cost-wise advantageous to consider the descendants of this node, i.e., the corresponding cluster/sub-network is indivisible. For further clarification see the illustrative example below.

**Preposition**: The set $P^{(opt)}$ of all optimal sub-networks represents the *optimal partitioning*.

## 12. ILLUSTRATIVE EXAMPLE

*Remark*7: For the sake of simplicity, we assume that the following sub-networks are not further divisible:
- $N_{10}$, $N_{11}$, $N_{33}$, $N_{39}$, $N_{58}$, $N_{63}$, $N_{77}$ {for instance, as initial conditions specified by a network designer};
- $N_6$, $N_{17}$, $N_{18}$, $N_{28}$, $N_{30}$, $N_{32}$, $N_{59}$, $N_{62}$, $N_{76}$ {for example, because they do not satisfy at least one of the threshold conditions}.

This example is presented with different forms of data handling: including a table, a binary tree and the corresponding arrays.

### 12.1 Computation of final labels
In the following Tables 4.1-4.3 we treat all indivisible subnetworks as leaves of the binary tree and indicate this with an underline.
From Tables 4.1-4.3 we determine:
- The set of all nodes $N_k$, for which $w_k := 1$; {totally *twenty* nodes for *k*=5; 6; 10; 11; 14; 16; 17; 18; 19; 28; 30; 32; 33; 39; 58; 59; 62; 63; 76; 77};
- The set of all *optimal* nodes $N_k^o$ {totally *ten* optimal nodes for *k*=5; 6; 14; 16; 17; 18; 19; 30; 62; 63}.

*NB*: In the Tables 4.1-4.3 the final labels, for which hold $L_j = F_j$, are shown in **bold** *italics*.

**Table 4.1**

| Subnetworks $N_k$ | $N_1$ | $N_2$ | $N_3$ | $N_4$ | $N_5$ | $N_6$ | $N_7$ | $N_8$ | $N_9$ | $N_{10}$ | $N_{11}$ |
|---|---|---|---|---|---|---|---|---|---|---|---|
| $t_k$ | 235 | 101 | 120 | 49 | 41 | 40 | 65 | 22 | 28 | 20 | 20 |
| $q_k$ | 35 | 29 | 23 | 25 | 14 | 20 | 21 | 13 | 12 | 10 | 13 |
| $L_k=d_k$ | 270 | 130 | 143 | 74 | 55 | 60 | 86 | 35 | 40 | 30 | 33 |
| $F_k$ | 248 | 119 | 129 | 64 | **55** | **60** | 69 | 32 | 32 | 30 | 33 |

**Table 4.2**

| $N_k$ | $N_{14}$ | $N_{15}$ | $N_{16}$ | $N_{17}$ | $N_{18}$ | $N_{19}$ | $N_{28}$ | $N_{29}$ | $N_{30}$ | $N_{31}$ | $N_{32}$ |
|---|---|---|---|---|---|---|---|---|---|---|---|
| $t_k$ | 24 | 26 | 9 | 10 | 11 | 8 | 10 | 12 | 10 | 12 | 4 |
| $q_k$ | 12 | 12 | 6 | 7 | 8 | 5 | 8 | 9 | 6 | 7 | 3 |
| $L_k=d_k$ | 36 | 38 | 15 | 17 | 19 | 13 | 18 | 21 | 16 | 19 | 7 |
| $F_k$ | **36** | 33 | **15** | **17** | **19** | **13** | 18 | 19 | **16** | 17 | 7 |

**Table 4.3**

| $N_k$ | $N_{33}$ | $N_{38}$ | $N_{39}$ | $N_{58}$ | $N_{59}$ | $N_{62}$ | $N_{63}$ | $N_{76}$ | $N_{77}$ |
|---|---|---|---|---|---|---|---|---|---|
| $t_k$ | 4 | 4 | 3 | 3 | 5 | 4 | 5 | 1 | 2 |
| $q_k$ | 5 | 2 | 4 | 5 | 6 | 3 | 5 | 1 | 1 |
| $L_k=d_k$ | 9 | 6 | 8 | 8 | 11 | 7 | 10 | 2 | 3 |
| $F_k$ | 9 | 5 | 8 | 8 | 11 | **7** | **10** | 2 | 3 |

As it follows from Tables 4.1-4.3, the minimal total cost of all hardware elements {the transmission links plus the routers/ESs} equals

$$F_5 + F_6 + F_{14} + F_{16} + F_{17} + F_{18} + F_{19} +$$
$$F_{30} + F_{62} + F_{63} = 55 + 60 + 36 + 15 \quad (12.1)$$
$$+17 + 19 + 13 + 7 + 10 = 248 = F_1.$$

**12.2 Search for optimal clusters via binary-tree algorithm**

Each node of the binary tree is described in form $\{k, w_k\}$, where computation of $w_k$ is described in (11.4).
As a result, we have the following list:

$$\{1,0\};\{2,0\};\{3,0\};\{4,0\};\{5,1\};\{6,1\};\{7,0\};$$
$$\{8,0\};\{9,0\};\{10,1\};\{11,1\};\{14,1\};\{15,0\};$$
$$\{16,1\};\{17,1\};\{18,1\};\{19,1\};\{28,1\};\{29,0\};$$
$$\{30,1\};\{31,0\};\{32,1\};\{33,1\};\{38,0\};\{39,1\};$$
$$\{58,1\};\{59,1\};\{62,1\};\{63,1\};\{76,1\};\{77,1\}.$$

Since for all ancestors of node $N_5$ hold $w_{a(5)} := 0$, therefore by the principle of optimality $N_5^o$ is optimal. Analogously, for all ancestors of the node $N_6$ hold $w_{a(6)} := 0$, therefore by the principle of optimality $N_6^o$ is also optimal.
Applying the principle of optimality we find that for $k=5; 6; 14; 16; 17; 18; 19; 30; 62; 63$ the nodes $N_k^o$ are indivisible, therefore they are *optimal*.
This algorithm is designed by the author of this paper.

## 13. STATISTICAL PROPERTIES OF COST-FUNCTION $h(x)$

More than *eighteen hundred* computer experiments confirmed that the cost-function $h(x)$ has rather stable statistical properties.
Let $R(x):=\left[\max h(x) - \min h(x)\right]/\min h(x)$ be the range of $h(x)$.
***Property*1**: if $n >> 10$ and $h(x)$ has a range $R(x)$ larger than 5%, then $h(x)$ is a *bimodal* function on the period $x \in (0, \pi)$;

*Property*2: if the range *R(x)* is smaller than 5% or the number of users is small (*n*<25), then *h(x)* has more than one local minimum. In this case, if the range *R(x)* is small, then *h(x)* is a shallow function and its optimization does not provide a substantial gain. On the other hand, if *n* is small, then time complexity to check all *n* rotations is also small. These statistical properties of the function *h(x)* have been discovered by the author of this paper twenty four years ago via numerous computer experiments.

Therefore, Property1 can be used to design a more elaborate algorithm that requires substantially less computation than the thorough parametric search over interval $x \in (0, \pi)$, [22], [23].

## 14. COMPLEXITY ANALYSIS OF OPTIMAL SEARCH FOR LARGE *n*

It is easy to see that the parametric partitioning requires in the general case exactly *n* rotations of the separating line *L*. As a result, the time-complexity *T(n)* to divide *n* users into *two* clusters equals $T(n) = an^2 + O(n)$ for large *n*. However, in the instances where *Property* 1 of *h(x)* is applicable, this complexity can be substantially reduced. In this case the search algorithm for a minimum of function *h(x)* requires O(log*n*) rotations of the separating line *L*. As a result, $T(n) = bn\log n + O(n)$ for large *n* and overall worst-case complexity is of order $O(n^2 \log n)$.

The methods of complexity evaluation developed by the author of this paper in [24] demonstrate that the *average* complexity of the overall binary partitioning is of order $O(n \log^2 n)$.

## 15. CONCLUSION

Several algorithms developed by the author are described in the paper. These algorithms provide a foundation for optimal design of configuration of terrestrial networks based on satellite communication. The author reduced the complexity of the problem by employing the statistical properties of the optimized function. It is demonstrated that the entire process of optimal design for large number of users has polynomial time complexity.

## ACKNOWLEDGEMENTS

I express my appreciation to E. Blum, P. Fay, R. Morrell, N. Pallav and A. Zhang and to anonymous referees for their comments and suggestions that improved the style of this paper.

# APPENDIX

## VALIDATION OF CoG ALGORITHM

Let

$$T_1(u,v) := \left(\sum_i w_i x_i / R_i\right) / \left(\sum_i w_i / R_i\right);$$
$$T_2(u,v) := \left(\sum_i w_i y_i / R_i\right) / \left(\sum_i w_i / R_i\right);$$
(A1)

where $R_i := \sqrt{(u-a_i)^2 + (v-b_i)^2}$. (A2)

**Lemma1:** $(u_*, v_*)$ is sufficiently close to location of a user $P_j$, then the iterative process, described in Steps B2-B3, terminates. In other words, for every $j=1,2,..,n$

$$\lim_{R_j \to 0} (T_1(u_*, v_*), T_2(u_*, v_*)) = (u_*, v_*).$$

*Proof*: Observe that

$$T_1(u,v) = \left(w_j x_j + R_j \sum_{i \neq j} w_i x_i / R_i\right) / \left(w_j + R_j \sum_{i \neq j} w_i / R_i\right).$$

Therefore $T_1(u,v) = x_j$ if $R_j = 0$.

Analogously, we deduce that $T_2(u,v) = y_j$ if $R_j = 0$.

**Lemma2:**
If $(u_0, v_0) = (T_1(u_0, v_0), T_2(u_0, v_0))$ (A3)

and for every $j=1,2,..,n$ holds $P_j \neq (u_0, v_0)$, then the stationary point $(u_0, v_0)$ is the optimal location of the router/ES.

*Proof*: In order to find the best location $(u, v)$ of router/ES we need to find minimum of function

$$F(u,v) = \sum_{i=1}^{n} w_i R_i(u,v). \quad (A4)$$

The process provided in Steps B1-B3 describes iterative solution of a system of two non-linear equations

$\partial F(u,v)/\partial x = 0$ and $\partial F(u,v)/\partial y = 0$. (A5)

These are necessary and sufficient conditions that $(u, v)$ is the optimal location of a router.

**Lemma3:** Let $F(u_0, v_0) = \min F(u,v)$,

and $(a_j, b_j) = (u_0, v_0)$.

Then $F(u_0, v_0) = \min[F(u,v) + w_j R_j]$. (A6)

*Proof* immediately follows from observation that $R_j(u_0, v_0) = 0$. (A7)

## ILLUSTRATIVE EXAMPLE: ASSOCIATED BINARY TREE

In the Fig.1-3 provided below are used the following legends:

1. The entire tree is represented as triad of sub-trees with roots $N_1, N_4, N_7$;
2. Each node $N_k$ with exception of leaves has two children $N_{2k}$ and $N_{2k+1}$;
3. Each node $N_c$ (with exception of $N_1$) has a parent $N_{\lfloor c/2 \rfloor}$;
4. A nodes framed in rectangle is a leaf of the tree; {leaves do not have children};
5. All other nodes are framed in ellipse; {such nodes have children/descendents};
6. Each node is described in the following format $\{L_k, N_k, F_k\}$.

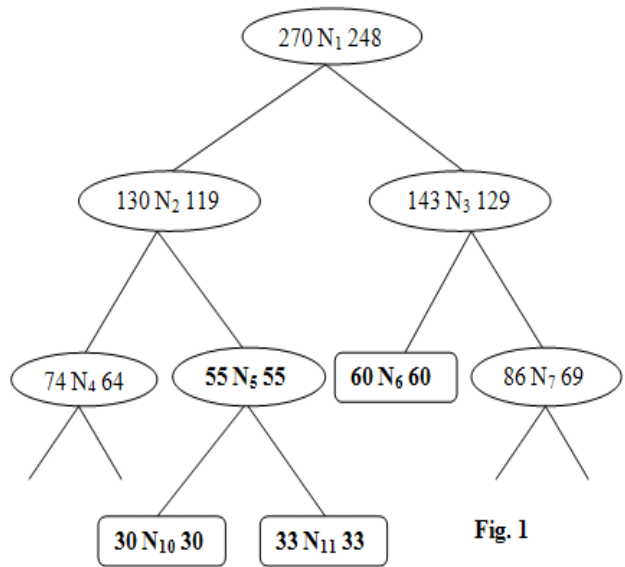

Fig. 1

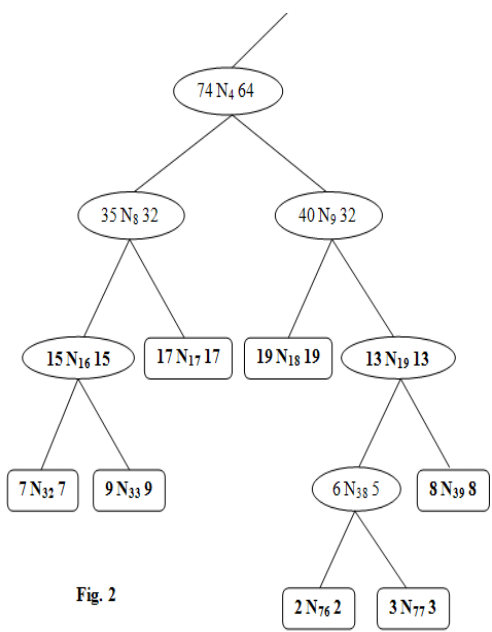

Fig. 2

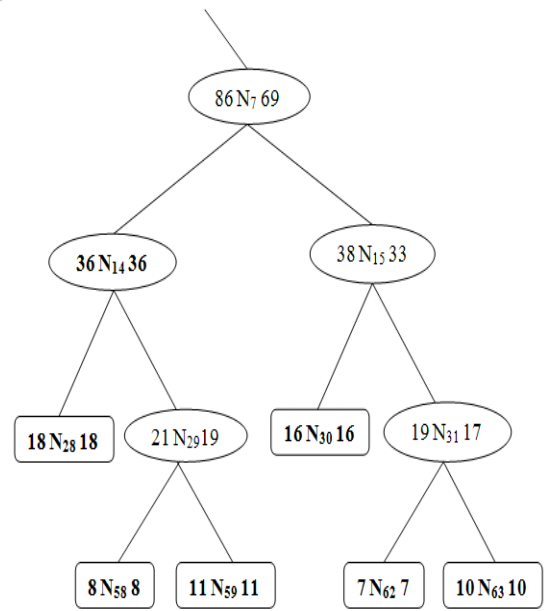

Fig. 3

**Boris S. Verkhovsky** is a Professor in the Computer Science Department at the New Jersey Institute of Technology. He received his PhD in computer science jointly from Latvia State University and from the Academy of Sciences of the USSR (Central Institute of Mathematics and Economics, Moscow). From his prior affiliations at the Scientific Research Institute of Computers (Moscow), Princeton University, IBM Thomas J. Watson Research Center, Bell Laboratories, University of Colorado and since 1986 at the NJIT, he acquired research interests and experience in cryptography, information assurance, communication security, optimal design and control of telecommunication systems, optimization of large-scale systems and algorithms design. For the last twenty years his research activity has been centered on design and analysis of cryptographic systems and information assurance algorithms. Professor Verkhovsky has published more than two hundred research papers. He is listed in Marquis *Who's Who* in America